\documentstyle{mn}
\title[Dust in 8C\,1435+635]
      {Detection of dust in the most distant known radiogalaxy}
\author[R.\ J.\ Ivison]
       {R.\ J.\ Ivison\\
        Royal Observatory, Blackford Hill, Edinburgh EH9 3HJ}
\date{Accepted 1995 May ??.
      Received 1995 May ??;
      in original form 1995 March 1}
\pagerange{000--000}

\begin{document}

\maketitle

\begin{abstract}
A search for millimetric continuum emission from eight
optically-selected, radio-quiet quasars and a radiogalaxy with $3.7 <
z < 4.3$, has been undertaken using a highly sensitive 7-channel
bolometer on the IRAM 30-m Millimetre Radio Telescope.  Detections of
a potentially dust-rich quasar, and of 8C1435+635, the most distant
known radiogalaxy, are reported. An extrapolation of the steepening
centimetric radio spectrum of 8C1435+635 accounts for less than one
per cent of the observed 1.25-mm flux density, indicating that the
emission is most likely from warm dust, although the present data
cannot discriminate against synchrotron emission. If the emission is
thermal, then the derived dust mass lies in the range, $2 \times 10^9
< M_{\rm d} < 8 \times 10^7$\,M$_{\odot}$ for $20 < T_{\rm d} <
100$\,K, or $M_{\rm d} \sim 1.6 \times 10^8$\,M$_{\odot}$ for $T_{\rm
d} = 60$\,K, similar to that derived for 4C41.17, suggesting a
molecular gas mass of between $4 \times 10^{10}$ and $9 \times
10^{11}$\,M$_{\odot}$. The quasar, PC2047+0123 at $z=3.80$, has no
detectable centimetric emission and the 1.25-mm continuum detected
here probably also originates from $1.5 \times 10^8$\,M$_{\odot}$ of
dust (again for $T_{\rm d} = 60$\,K). Upper limits have been obtained
for four quasars, corresponding to dust mass limits of around
$3\,\sigma < 2 \times 10^8$\,M$_{\odot}$; less useful limits have been
set for a further three quasars.
\end{abstract}

\section{Introduction}

 \begin{table*}
 \caption{1.25-mm photometry.}
 \begin{tabular}{lccccccrl}
Source             &Object&$z$  &\multicolumn{2}{c}{Adopted coordinates
(B1950)}                                                        &$\lambda_0$&UT
Date&
Flux Density       &Note\\
name               &type  &     &$\alpha$   &$\delta$   &/$\mu$m    &(1995
Feb)&($\pm \sigma$) /mJy&/Reference\\
&&&&&&&&\\
PC0027+0521        &RQQ   &4.21 &00 27 30.10&+05 21 39.0&240        &20.6
&$-0.27 \pm 4.66$   &1, SSG94\\
BRI0151$-$0025     &RQQ   &4.20 &01 51 05.90&$-$00 25 44.0&240      &20.7
&$0.71 \pm 2.25$    &1, S94\\
PC1301+4747        &RQQ   &4.00 &13 01 50.80&+47 47 35.0&250        &21.2
&$1.20 \pm 1.77$    &SSG91, SSG94, SVSG92\\
8C1435+635         &RG    &4.26 &14 35 27.50&+63 32 12.8&238        &17.3
&$3.03 \pm 0.85$    &2, 3, L94, S95\\
                   &      &     &           &           &           &19.3
&$2.61 \pm 0.70$    &3, 4\\
                   &      &     &           &           &           &22.2
&$2.26 \pm 0.67$    &3, 5\\
PC1640+4628        &RQQ   &3.70 &16 40 37.20&+46 28 01.0&266        &18.3
&$1.08 \pm 1.13$    &SSG91, SSG94, SVSG92\\
PC1643+4631A       &RQQ   &3.79 &16 43 33.50&+46 31 38.0&261        &18.5
&$0.11 \pm 1.00$    &6, SSG91, SSG94, SVSG92\\
PC1643+4631B       &RQQ   &3.83 &16 43 52.40&+46 31 02.0&259        &19.5
&$-0.95 \pm 0.81$   &SSG91, SVSG92\\
PC1644+4744        &RQQ   &3.70 &16 44 48.40&+47 44 24.0&266        &20.5
&$-0.03 \pm 0.95$   &SSG94\\
PC2047+0123        &RQQ   &3.80 &20 47 50.70&+01 23 56.0&260        &17.5
&$1.51 \pm 0.91$    &7, SSG91, SVSG92\\
                   &      &     &           &           &           &19.6
&$1.97 \pm 1.19$    &7\\
                   &      &     &           &           &           &22.4
&$2.11 \pm 0.64$    &5, 7\\
 \end{tabular}

 \vspace*{2mm} \noindent 1: high rms due to poor weather; 2: $\tau =
 0.31$; 3: weighted mean is $2.57 \pm 0.42$\,mJy (6.1\,$\sigma$); 4:
 $\tau = 0.12$; 5: $\tau = 0.08$; 6: 5-arcsec SW of SSG94 position; 7:
 weighted mean is $2.08 \pm 0.47$\,mJy (4.4\,$\sigma$); L94: Lacy et
 al.\ (1994); S94: Smith et al.\ (1994); S95: Spinrad et al.\ (1995);
 SSG91: Schneider, Schmidt \& Gunn (1991); SSG94: Schneider, Schmidt
 \& Gunn (1994); SVSG92: Schneider et al.\ (1992).  \end{table*}

There are approximately 50 known quasars and radiogalaxies (RGs) with
$z \sim 4$. It is natural to employ these objects as observational
probes of the early stages of galaxy formation; for example,
low-redshift RGs represent a relatively homogeneous group in many
respects, and comparison with RGs at higher redshifts gives some
perception of their evolutionary traits, always bearing in mind fears
about sample bias, e.g.\ the alignment effect (McCarthy et al.\ 1987;
Chambers, Miley \& Van Breugel 1987).

Young quasars and RGs are likely to be the sites of active star
formation, and they are therefore an obvious place to search for
thermal emission from dust and line emission from molecular gas
(McMahon et al.\ 1994).  Evidence already supports the idea that there
are vast reservoirs of dust-rich gas in several high-redshift objects,
notably in the radio-quiet quasar BR1202$-$0725 (Isaak et al.\ 1994;
McMahon et al.\ 1994), the Cloverleaf quasar (Barvainis et al.\ 1994),
and the RG 4C41.17 (Dunlop et al.\ 1994; Chini \& Krugel 1994), all of
which have been detected in the rest-frame far-infrared. A thermal
origin for the millimetric emission from these objects would indicate
dust masses of around $10^8$--$10^9$\,M$_{\odot}$, and it is hard to
explain their submillimetre spectral indices ($\alpha \sim 3$--$4$,
where $F_{\nu} \propto \nu^{\alpha}$) by anything {\em other} than
emission from dust (e.g.\ Hughes et al.\ 1993).

Determining the evolutionary phase of these objects (e.g.\ whether a
`proto-galactic' label applies) is not straightforward. One could
measure the fraction of their gas which has been converted into stars,
but this relies on knowledge of the relative gas-to-dust ratios over a
range of redshifts, which emphasizes the importance of both continuum
and molecular-line data. The millimetric continuum data presented here
will be complemented by centimetric spectral-line observations of CO
(Ivison et al., in preparation), which should yield estimates of
several important physical parameters, not least the mass and
temperature of the gas and dust.

\section{Bolometer array measurements}

The data were obtained during 1995 February 17--22 using the MPIfR
7-channel $^3$He-cooled bolometer array (see Kreysa 1993) and the 30-m
IRAM Millimetre Radio Telescope on Pico Veleta, Spain. The individual
beamsizes are 11--12\,arcsec (HPBW), and the bolometers are separated
by 22\,arcsec in a hexagonal arrangement surrounding a central
pixel. During our observing session we employed a 1-arcmin azimuthal
chop-throw, at a rate of 2\,Hz; the telescope was also
position-switched by 1\,arcmin every 10\,s in the standard symmetric
ON-OFF-OFF-ON mode. The nett result was that one third of the total
observing time was spent on the source.

Typically 15--20 samples were obtained for each of the targets
(separated by pointing, focusing, opacity tips and calibration scans
of Uranus and Mars, for which $T_{\rm b}$ of 101 and 205\,K were
assumed, respectively), each consisting of forty 10-s sub-samples. The
weighted mean of the outer channels was subtracted from the central
channel --- an effective method of reducing the effects of sky-noise,
i.e.\ the fraction of atmospheric emission that remains after
chopping. The samples were then concatonated, tested for spikes, and
corrected both for atmospheric attenuation and for a gain-elevation
dependence, $G_{\rm el}$, of the form:

\begin{equation}
G_{\rm el} \;\; = \;\; ({\rm cos}({\rm el}-45^{\circ})^2 + 0.4 \;
({\rm sin}({\rm el}-45^{\circ})^2)^{-1} \;\;\; .
\end{equation}

The sky transparency at 1.25\,mm was virtually constant during each
session; it varied from night to night, but was generally good, with
the opacity ranging from 0.06 to 0.44. All the sources were observed
at low zenith distances ($<45^{\circ}$). The pointing characteristics
of the telescope were excellent, with rms fluctuations at 20 per cent
of a single-channel HPBW.

The $z=4.26$ RG, 8C1435+635, and the $z=3.80$ quasar, PC2047+0123,
were each observed on three separate occasions in order to confirm
that their emission was real. The flux-density measurements are
presented in Table~1, together with redshifts, rest wavelengths and
adopted coordinates.

\section{Results and discussion}

 \begin{figure*} \vspace{12cm} \caption{Spectral energy distribution
of 8C1435+635. The line drawn in the rest-frame far-IR illustrates a
60-K greybody, with +2 frequency dependence for the dust-grain
emissivity (neither parameter is reliably constrained).}
\end{figure*}

The data from Table~1 reveal that 8C1435+635, a $z=4.26$ RG (Lacy et
al.\ 1994, hereafter L94; Spinrad et al.\ 1995, hereafter S95), was
detected at the 6.1-$\sigma$ level, and the radio-quiet quasar,
PC2047+0123 at $z=3.80$, was marginally (4.4\,$\sigma$) detected.
Limits of $3\,\sigma < 3$\,mJy have been set for four quasars,
together with higher limits for a further three.

\subsection{8C1435+635}

The integrated spectrum of 8C1435+635 steepens from $\alpha=-1.2$ to
$-2.2$ via $-1.5$ between 0.151, 1.5, 4.9 and 8.4\,GHz; it then
continues at $\alpha=-2.2$ to 15\,GHz.  8C1435+635 was initially
selected for blind spectroscopic observations by L94 on the basis of
its steepening radio spectrum.

Fig.~1 shows the measurement obtained here for 8C1435+635, and those
obtained by L94 and S95 at lower and higher frequencies. Flux
densities upper limits from the {\em Infrared Astronomical Satellite}
have also been estimated at 12, 25, 60 and 100\,$\mu$m by searching a
1 square degree field centred on the RG for sources from the {\em
Faint Source Catalogue} ({\em FSC}), adopting the faintest in each
band as the upper limit. This crude method relies on the fact that if
the {\em FSC}'s sophisticated search routines cannot find a point
source, then the source must be below the 3-$\sigma$ threshold. The
method is probably as reliable as any other, and is less prone than
some to providing misleadingly low limits.

If the cm-wave radio spectrum continues to decline above 15.2\,GHz
with $\alpha=-2.2$ then the non-thermal contribution at 243\,GHz would
be 15\,$\mu$Jy. Between 15 and 243\,GHz, $\alpha=-0.3$, making it a
safe bet that its contribution to the 243-GHz flux density is
completely insignificant, and that the 243-GHz emission is probably
dominated by thermal emission from dust. Of course, it is not possible
to discriminate against a synchrotron origin, and absolute
confirmation of the emission mechanism must be obtained at frequencies
above or below 243\,GHz; the most reliable method would be to obtain
photometry at 0.8\,mm using UKT14 on the 15-m James Clerk Maxwell
Telescope, where the thermal emission contribution would be around
15\,mJy for $\alpha=+4.0$ (contributions of +2.0 from both the dust
emissivity and the Rayleigh-Jeans law) --- within the capabilities of
the instrument on an excellent night (e.g.\ Dunlop et al.\ 1994).

If the 243-GHz emission mechanism is thermal, and assuming a range of
estimates for the dust temperature based on the $z=3.8$ RG, 4C41.17,
and on samples of low-redshift quasars and RGs (Knapp \& Patten 1991;
Andreani, La Franca \& Cristiani 1993; Chini \& Kr\"{u}gel 1994;
Dunlop et al.\ 1994), i.e.\ $T_{\rm d} = 20$--$100$\,K, and an
Einstein - de Sitter universe where
$H_0=50$\,km\,s$^{-1}$\,Mpc$^{-1}$, then for optically thin emission
we can determine the dust mass using (from Chini \& Kr\"{u}gel 1994):

\begin{equation}
M_{\rm d} \;\; = \;\; \frac{F_{\nu_{\rm obs}} \; D_{\rm L}^2}{(1+z) \;
\kappa_{\nu_0} \; B_{\nu_0}(T_{\rm d})} \;\;\; ,
\end{equation}

\noindent
where $F_{\nu_{\rm obs}}$ is the observed flux density, $\nu_{\rm
obs}$ and $\nu_0$ are the observed and rest-frame frequencies,
$\kappa_{\nu_0}$ is the dust absorption coefficient
(10.5\,cm$^2$\,g$^{-1}$ has been adopted here, from Kr\"{u}gel, Steppe
\& Chini 1990), and the luminosity distance is given by

\begin{equation}
D_{\rm L} \;\; = \;\; \frac{2 \; c \; (1 + z - (1 + z)^{0.5})}{H_0} \;\;\; .
\end{equation}

For $T_{\rm d}=60$\,K, the resulting dust mass is $1.6 \times
10^8$\,M$_{\odot}$, which is similar to that calculated for 4C41.17
using the same method. For $20 < T_{\rm d} < 100$\,K, $2 \times 10^9 <
M_{\rm d} < 8 \times 10^7$\,M$_{\odot}$. Note that the above
calculation is sensitive to parameters assumed for both the dust and
the cosmology. Conversion from dust mass to gas mass is troublesome
even in the Milky Way; but adopting $M_{\rm g}/M_{\rm d} = 500$ (for a
system which probably has a relatively low abundance of heavy
elements) gives a gas mass, $M_{\rm g}$, of $8 \times
10^{10}$\,M$_{\odot}$ for $T_{\rm d}=60$\,K.

Dust-rich gas in these quantities may explain the anomalously weak
Lyman $\alpha$ luminosity of this RG, as suggested by L94. Following
Chini \& Kr\"{u}gel (1994), in the idealized case where
$10^8$\,M$_{\odot}$ of dust is distributed in a 50-kpc diameter
sphere, the optical depth at 1216\,\AA\ would be 1--2, sufficient to
entirely absorb the Lyman $\alpha$ photons assuming that neutral
hydrogen is sufficiently abundant to make resonant scattering
important (where scattering increases the path length and,
accordingly, the optical depth to absorption by dust; see Eales \&
Rawlings 1993). This does not, however, explain why the Lyman $\alpha$
luminosity of 8C1435+635 is low relative to 4C41.17, another distant,
dust-rich radiogalaxy. S95 note that 8C1435+635 lies within the
scatter of the Lyman $\alpha$ luminosity---radio power relationship.

\subsection{Quasars}

We observed a total of eight radio-quiet quasars, all of which satisfy
$3.7 < z < 4.3$. One was detected, PC2047+0123 at $z=3.80$, with a
signal-to-noise ratio of $4.4$ after taking the weighted mean of data
obtained on separate nights.  There is nothing intrinsically flawed
about this procedure, but we regard the detection as slightly
marginal, and confirmation should be sought with one of the upcoming
generation of submillimetre bolometer arrays. The rms flux densities
for the remaining seven quasars (Table~1) yield 3-$\sigma$ upper
limits of between $10^8$ and $10^9$\,M$_{\odot}$ for their dust
masses, using the same assumptions as for 8C1435+635.

Upper limits ($3\sigma$) of $225$ and $120$\,$\mu$Jy were obtained for
PC2047+0123 at 1.5 and 4.9\,GHz by Schmidt et al.\ (1995) and
Schneider et al.\ (1992), respectively; yielding a lower limit for the
spectral index between 4.9 and 243\,GHz of $\alpha>+0.7$. This
provides evidence, always bearing in mind the marginal nature of the
detection and the lack of a measurement of the submillimetre spectral
index, that the 1.25-mm flux density is dominated by thermal emission
from dust, though again we cannot discriminate against a non-thermal
mechanism. If dust is responsible, we estimate $M_{\rm d} = 1.5 \times
10^8$\,M$_{\odot}$ (under the assumptions discussed in the previous
section, and with $\kappa_{\nu_0} = 8.7$\,cm$^2$\,g$^{-1}$).

\section{Concluding remarks}

In an attempt to increase the number of known gas-rich systems at $z
\sim 4$, eight radio-quiet quasars and a RG, which satisfy $3.7 < z <
4.3$, have been observed using a sensitive 7-channel bolometer on the
30-m IRAM Millimetre Radio Telescope, resulting in the detection of
1.25-mm continuum emission from the radio-quiet quasar, PC2047+0123 at
$z=3.8$, and from 8C1435+635, the most distant known RG. Both are now
prime targets for spectral-line observations of molecular gas. We have
placed limits of $3\,\sigma < 3$\,mJy on a further four high-redshift
quasars, which corresponds to approximately $3\,\sigma < 2 \times
10^8$\,M$_{\odot}$ of dust (for $T_{\rm d} \sim 60$\,K; $q_0=0.5$;
$H_0=50$\,km\,s$^{-1}$\,Mpc$^{-1}$).  Another three quasars have
slightly less useful limits.

If the 1.25-mm emission from 8C1435+635 is dominated by thermal
emission, which seems likely given the steep centimetric spectrum,
then the derived dust mass is similar to those found in nearby
radio-quiet quasars (Hughes et al.\ 1993), and 1--2 orders of
magnitude higher than those found for nearby radio galaxies
(e.g. Knapp, Bies \& van Gorkom 1990; Knapp \& Patten 1991). There is
evidence, therefore, that spectral-line and sub-mm continuum
observations of this RG will provide clues concerning the dynamical,
physical and evolutionary state of some of the most ancient known
material in the Universe, and there is a tantalizing possibility that
the Very Large Array can be used to map the molecular gas with the
same angular resolution as can be routinely obtained when observing CO
in local galaxies (0.1\,arcsec $\sim$ 1\,kpc at $z \sim 4$, for
$\Omega_0 = 0.5$ and $H_0 = 50$\,km\,s$^{-1}$\,Mpc$^{-1}$) .

\subsection*{ACKNOWLEDGMENTS}

It is a pleasure to acknowledge the excellent support of Raphael
Moreno at Pico Veleta.

\bsp

\end{document}